\documentclass[runningheads]{llncs}
\usepackage{graphicx}
\usepackage{booktabs}
\usepackage{multirow}
\usepackage[utf8]{inputenc} 

\begin{document}
\title{The Explanatory Gap in \\Algorithmic News Curation}
\titlerunning{Explanatory Gap in ANC}
%
%
%

\author{Hendrik Heuer\orcidID{0000-0003-1919-9016}}
\authorrunning{Heuer}
%
\institute{
University of Bremen, Institute for Information Management (ifib) \& Centre for Media, Communication and Information Research (ZeMKI), Germany\\
\email{hheuer@uni-bremen.de}
}

\maketitle              
\begin{abstract}
Considering the large amount of available content, social media platforms increasingly employ machine learning (ML) systems to curate news. This paper examines how well different explanations help expert users understand why certain news stories are recommended to them. The expert users were journalists, who are trained to judge the relevance of news. Surprisingly, none of the explanations are perceived as helpful. Our investigation provides a first indication of a gap between what is available to explain ML-based curation systems and what users need to understand such systems. We call this the Explanatory Gap in Machine Learning-based Curation Systems.

\end{abstract}
\section{Introduction}

Machine learning~(ML)-based curation systems are frequently applied to suggest products, restaurants, movies, songs, and other content. Such systems have become a ubiquitous part of users' daily experience of information systems~\cite{Jannach:2016:RSB:3013530.2891406}. On social media sites like Facebook and Twitter, ML-based curation systems solve the challenging tasks of selecting, organizing, and presenting news from a variety of sources~\cite{eslami_i_2015}. While curation is necessary considering the large number of users of social media sites and the immense number of available news stories, ML-based curation systems pose important challenges regarding algorithmic transparency and algorithmic experience~\cite{doi:10.1080/21670811.2016.1208053,rader_explanations_2018,alvarado_towards_2018,doi:10.1177/20539517211017593}. In the past, news curation was a task predominantly performed by skilled journalists, who assessed the newsworthiness of content~\cite{Trielli:2019:SNC:3290605.3300683}. Increasingly, this task is performed by complex and opaque algorithms that lack transparency. This is problematic since social media platforms, which rely on ML-based curation systems, are becoming an important source of news~\cite{foundation_american_2018,DBLP:journals/corr/abs-1201-4145,gottfried_news_2016}. Two-thirds of 18-24 year-olds worldwide rely on social media for news~\cite{newman_reuters_2019}. Facebook's News Feed is the canonical example of an ML-based curation system that is used daily by a large number of users. A large majority of U.S. adults using Facebook's News Feed thinks they have little~(57\%) or no control~(28\%) over the news curation system~\cite{pewresearch_many_2019}. More than half of the respondents also said they do not understand why certain posts are included by the ML-based curation system. Only every seventh person~(14\%) thinks that they understand the curation on Facebook very well.

This paper explores how the simplicity, intuitiveness, and interactivity of explanations influences users' understanding of personalized recommender systems for news. Despite the active research on adaptation and personalization, little is known about how to best implement explanations for such systems and how such explanations are perceived by users~\cite{Millecamp:2019:EEE:3301275.3302313}. While researchers try to take aspects like novelty, diversity, unexpectedness, and utility into account for the evaluation of recommendation systems~\cite{Jannach:2016:RSB:3013530.2891406}, a research gap exists regarding the understanding of explanations for personalized recommender systems. With this paper, we address this research gap and conduct a user study where expert users use an ML-based curation system. The system provides three types of ML explanations that we selected based on the design criteria simplicity, intuitiveness, and interactivity~\cite{DixFinlayEtAl03,shneiderman_2004}.

We conducted a user study with 25 professional journalists who trained personalized curation system by rating news stories in blocks. The ML-based curation system included the following explanations: (1)~system predictions grouped by the confusion matrix~(intuitiveness), (2)~performance metrics like accuracy, precision, and recall commonly used to evaluate machine learning systems~(simplicity), and (3)~an interactive ranking of the most important keywords according to the curation system~(interactivity). Users were able to interact with the (3) ranking of keywords by changing the importance of individual words which changed the feature importance in the model. Participants used all three explanations six times. After reviewing the recommendations and explanations with varying levels of system performance, participants rated how well the explanations supported their understanding of the curation system and how helpful they found the explanations. We also compare their understanding of the curation system to how well they think they understand Facebook's News Feed. Our analysis provides a first indication of an explanatory gap between what is available to explain curation systems and what users need to understand such systems. This gap exists for all three explanations, regardless of whether they are designed to be simple, intuitive, or interactive.

\section{Background}

Adaptive systems for news personalization have a long history~\cite{ardissono2001adaptive,info10100300,Sheidin:2017:VSE:3030024.3040984}. Facebook, as one of the most widely used ML-based curation systems, cites three signals that are used to predict and rank the relevance of the content: what kind of content it is, who posted it, and how users interact with the content~\cite{facebook_newsfeed_2018}. In our investigation, we focus on the basic specialization use case of selecting news, i.e. we do not take postings from other users into account. Our research connects to Hamilton et al., who highlight the importance of studying where, when, and how users are made aware of algorithms and how the perception translates into knowledge about the process at hand~\cite{hamilton_path_2014}. Amershi et al. argue that explicitly studying the users of learning systems is critical to advancing the field~\cite{amershi_power_2014}. This connects to a large body of research on explanations that are derived in specific contexts, but whose helpfulness is not evaluated in experimental user studies~\cite{Ribeiro_2016_WIT,DBLP:journals/corr/StrobeltGHPR16}. Konstan and Riedl identified the most important open research problems and key challenges of recommender systems. They argue that the user experience of such systems needs more attention~\cite{Konstan2012}. For Konstan and Riedl, the user experience is the delivery of the recommendations to the user and the interaction of the user with those recommendations. This view is supported by Jugovac and Jannach, who found that a large body of research is focused on the problem of rating prediction and item ranking while other aspects receive comparatively little attention~\cite{Jugovac:2017:IRR:3143523.3001837}. This paper focuses on the classification of news, not the ranking of news or the prediction of ratings.

In the context of ML-based curation systems, transparency is especially important since research showed that it positively influences users' trust in systems~\cite{Jannach:2016:RSB:3013530.2891406}. Eiband et al. analyzed 35,000+ app store reviews of three popular Android apps regarding interaction problems that can be attributed to algorithmic decision-making~\cite{Eiband:2019:PAM:3301275.3302262}. They investigate user reviews of the mobile applications of Facebook and Netflix, which both rely on ML-based curation systems. Their analysis shows how timely the call for more transparency and better explanations of curation systems is. Eiband et al. highlight the importance of user control and explanations of output. They identified problems with the curation algorithm, e.g. the biases enacted by the algorithm and the way the algorithm ranked the results. They also found that users want more control over their feed. Overall, the investigation highlights the importance of intuitive, simple, and interactive explanations, which motivated this research. 

Despite a large consensus that explanations are helpful and that algorithmic transparency is important~\cite{Tintarev2012,Geiger2017,doi:10.1080/21670811.2016.1208053}, the amount of empirical research that investigates explanations of curation systems in experimental user studies is limited, with a few notable exceptions focused on Facebook~\cite{rader_understanding_2015,rader_explanations_2018} and YouTube~\cite{10.1145/3415192,10.1145/3473856.3473864}. Furthermore, McNee et al. found that user satisfaction does not always correlate with high recommender accuracy~\cite{McNee:2006:AEA:1125451.1125659}. They show that the evaluation of such systems can be classified as the similarity of recommendation lists, recommendation serendipity, and the importance of user needs and expectations in a recommendation~\cite{McNee:2006:AEA:1125451.1125659}. Experimental studies in specific contexts are crucial because the context of recommender systems is known to shape the evaluation criteria of users~\cite{Jannach:2016:RSB:3013530.2891406}. We, therefore, focus on news recommendations. Prior research showed that the task of providing explanations for an ML-based curation system is difficult. Green et al. found that insufficient research has considered how the interactions between people and models influence human decisions~\cite{Green:2019:PLA:3371885.3359152}. This is especially important for news, which directly influence how people perceive the world and which can potentially affect their political opinions. Rader et al. investigated how explanations can support algorithmic transparency in the context of Facebook's News Feed~\cite{rader_explanations_2018}. They explored different explanation styles ranging from black-box scenarios describing the motivation of a system over white box scenarios that describing inputs and outputs of a system or how the system works. They found that all explanations made participants more aware of how the system works and helped them detect biases. At the same time, the explanations were not helping participants evaluate the correctness of the system’s output, which directly informed our research questions about whether explanations improve expert users' understanding of the quality of ML-based curation systems. Their research motivated us to focus on explanations of the model as a whole and to design novel explanations that go beyond the different explanation styles they explored.

\section{Method}

We designed three explanations based on the design criteria simplicity, intuitiveness, and interactivity regarding their helpfulness in the context of ML-based curation systems. These explanations make it transparent to users how well the system they are interacting with performs and how well the recommendations of a system are personalized to the user. This study addresses the following research questions:

\begin{itemize}
\item Do explanations focused on simplicity, intuitiveness, and interactivity improve expert users' understanding of an ML-based curation system (RQ1)? 
\item Which of the explanations is perceived as the most helpful in understanding news recommendations (RQ2)? 
\item How does the ability to change the curation system affect system performance (RQ3)?
\end{itemize}

To answer these research questions, we conducted an online study with professional journalists who trained personalized ML-based curation systems. The study consisted of two parts: rating news articles and evaluating curation systems. Before the study, participants were asked basic demographic questions regarding gender, age, and highest education. In the study, participants rated individual news articles using a Tinder-like swiping interface. The swiping interface was explained with a video. Participants rated six blocks of 12 news stories. After each block, a new machine learning model was trained. We trained the models with different amounts of training data, ranging from 10 to 60 news stories for each of the 25 users. The ML systems were trained with an 80\%-train-20\%-test-split so that the amount of test data to compute accuracy, precision, and recall was proportional to the amount of training data. For the sixth system, 60 news stories were used to train the system and 12 news stories were used to evaluate it. To compute reliable ML statistics, we performed 5-fold cross-validation~\cite{mueller2016introduction}.

Participants were presented with personalized predictions by the systems and three explanations based on design considerations explained in the following section. At the end of the experiment -- after having used the explanations six times -- participants rated the helpfulness of the three explanations on an 11-point Likert scale. Participants also rated how well they understood why certain posts are included by the system and others are not. The possible answers included \textit{``Not well at all''}, \textit{``Not very well''}, \textit{``Somewhat well''}, \textit{``Very well''}, and \textit{``Don't know''}. We compared this to how well the participants understood why certain posts are included in Facebook's News Feed, a widely used ML-based curation system that does not provide such explanations.

\subsection{Sampling and Participants}

Our sampling strategy was aimed at recruiting professional journalists who are an ideal target audience to compare different explanations of curation systems because journalists are familiar with the task of news curation. This connects to prior research with extreme users which showed that they can provide rich insights into issues like customization in communication apps and can be generalized to other users~\cite{griggio_customizations,Choe:2014:UQP:2556288.2557372,Djajadiningrat:2000:IRE:347642.347664}. Journalists are trained to judge what content is relevant and whether the content provided is balanced and fair. To recruit journalists, we identified newsletters of associations of journalism and communication science as well as online groups focused on journalism on a career-oriented social network. We also contacted local news outlets through their executive editors and their press spokespeople. On all channels, we published the same call for participation. Each participant had a chance to win one of ten 10€ vouchers or to have 10€ donated to charity. Seventy-seven percent of participants decided to donate their incentive to charity. Through this self-selection sampling, we recruited 25 professional journalists from Germany. The mean age of participants was 41.76 years with a standard deviation of 12.76. The youngest participant was 26, the oldest participant was 70. Thirteen participants identified as male (52\%), ten as female (40\%). Two chose not to disclose their gender. Our sample is highly educated. The large majority of participants (84\%) have a university degree. All participants had a high-school equivalent education. Regulatory requirements regarding the welfare, rights, and privacy of human subjects were followed.

\subsection{Explanations for ML-based Curation System}

\begin{figure}[t]
\centering
  \includegraphics[width=\columnwidth]{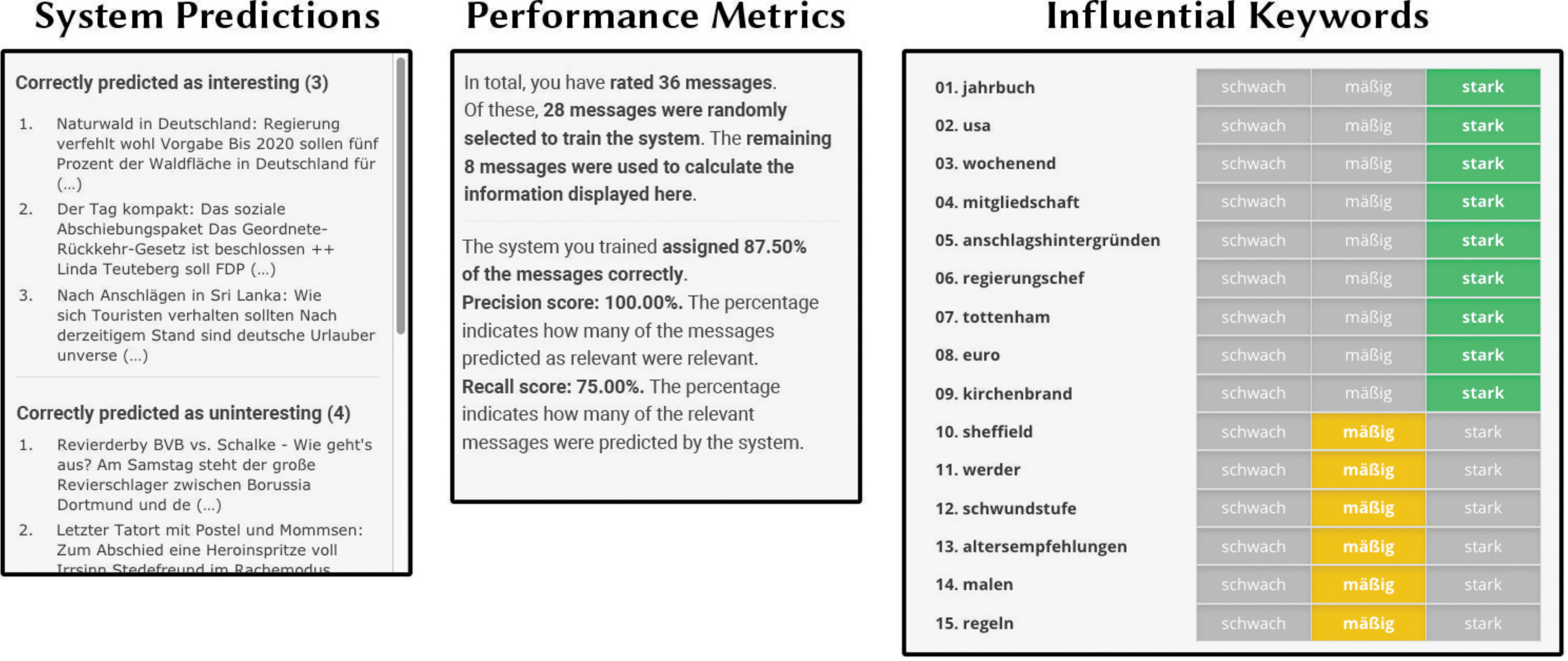}
  \caption{Three explanations were shown to journalists. 1.~System Predictions, i.e. predictions grouped by the confusion matrix, 2.~Performance Metrics like accuracy, precision, and recall. 3.~Influential Keywords and whether their influence on the model is weak, moderate, or strong.}~\label{fig:news_mixtures_types}
\end{figure}

In the study, each participant trained a personalized news curation system on a binary text classification task. The system was trained using the ratings that the user provided. Users interacted with the ML-based curation system through a web application. The task of the curation system was to predict whether a news story is interesting to a particular user or not. We developed the curation system from scratch to be able to change the ML model. The system predicts the interest in a story ($y$) given the nouns ($x_{i:n}$) in the story. We selected the Gaussian Na\"ive Bayes classifier as one of the most efficient and effective inductive learning algorithms for classification~\cite{Zhang2004,NIPS2001_2020}. The Gaussian Na\"ive Bayes classifier is a supervised ML algorithm that applies Bayes' theorem while assuming conditional independence between words~\cite{Maron:1961:AIE:321075.321084}. The Gaussian Na\"ive Bayes classifier is based on conditional probability, which makes the classifiers efficient to compute, straightforward to directly manipulate, and comparatively easy to explain. To train the curation system, participants were presented with a diverse mix of randomly selected news articles, political articles, cultural articles as well as articles about football. For this, we collected 413 recent news articles from the German public-service broadcaster (ARD) and the news magazine with the widest-circulation (DER SPIEGEL). Participants rated a subset of these articles. These ratings were then used to train the personalized curation systems. For both the rating and the training of the curation system, we used the nouns in the teaser of the article, which empirically provided sufficient information for the prediction task in our investigation.

In this study, we compare three explanations shown in Figure~\ref{fig:news_mixtures_types} that we designed based on the design criteria simplicity (\textbf{System Predictions}), intuitiveness (\textbf{Performance Metrics}), and interactivity (\textbf{Influential Keywords}). The \textbf{System Predictions} explanation presents participants with all predictions made by a personalized ML-based curation system. Participants were shown the headlines of all news from the test set in the four groups of the confusion matrix~\cite{mueller2016introduction}. These groups include true positives~($t_p$), true negatives~($t_n$), false positives~($f_p$), and false negatives~($f_n$). True positives ($t_p$) are interesting news stories that are correctly predicted as interesting news stories, true negatives ($t_n$) are uninteresting news stories correctly predicted as uninteresting. False positives ($f_p$) are uninteresting news stories that are predicted as interesting. False negatives ($f_n$) are interesting news falsely predicted as uninteresting. We included the system predictions as intuitive explanations because they present the predictions in a format that is similar to how news recommendations are encountered by users~\cite{powers2011evaluation,amershi_power_2014,mueller2016introduction}. We also presented the participants with the three most important \textbf{Performance Metrics} for ML systems: accuracy, precision, and recall~\cite{Goodfellow:2016:DL:3086952,10.1145/3340631.3394873}. Accuracy is defined as the percentage of correctly predicted news, i.e. $\frac{t_p + t_n}{t_p + t_n + f_p + f_n}$. Accuracy is one of the most widely used ML metrics in textbooks~\cite{Goodfellow:2016:DL:3086952,mueller2016introduction}. We also included precision as the proportion of the predicted news that is relevant~\cite{Rijsbergen:1979:IR:539927}: $\frac{t_p}{t_p + f_p}$. Recall is the proportion of interesting news covered by the predictions~\cite{Rijsbergen:1979:IR:539927}: $\frac{t_p}{t_p + f_n}$. The performance metrics were selected for their simplicity. Accuracy, precision, and recall all provide a single number that indicates the performance of a system, thus reducing the complexity of evaluating the quality of a system to a single, comparable number. Participants were also presented with the Top-15 most \textbf{Influential Keywords} of the Na\"ive Bayes classifier. The most influential keywords are the words with the highest prior probability for the class \textit{interesting}. To render the prior probabilities of the Na\"ive Bayes classifier more human-interpretable, we scaled the probabilities to values between 0 and 100. We classified the influence of a keyword on the prediction into the three categories weak, medium, and strong. Weak are keywords with a score smaller than 25. Medium keywords have a score between 25 and 50. Strong keywords have a score between 51 and 100. The thresholds were determined empirically based on the experience gained from training a large number of models. The Influential Keywords explanation was motivated by work on interactive machine learning and the explainability of machine learning~\cite{STUMPF2009639,kim2015interactive,DBLP:journals/corr/SelvarajuDVCPB16}. The approach is modeled after the feature importance that can be computed for decision trees~\cite{mueller2016introduction}. We implemented it as a Na\"ive Bayes classifier, which allowed us to directly manipulate the posterior probability of individual keywords. Since prior research shows that interactivity influences the user experience of ML systems~\cite{STUMPF2009639,tullio_how_2007,amershi_power_2014}, we also investigated how users interact with a curation system and how this affects system performance. Half of the participants were able to change the influence of the Top-15 keywords. Those with even IDs were able to change the influence of the keywords, those with odd IDs were not able to change the influence.

\section{Results}

We presented expert users with the three explanations shown in Figure~\ref{fig:news_mixtures_types} and studied whether the three explanations support them in understanding the news recommendations they receive. The large majority~(60\%) of participants stated that their understanding of why news stories were included by the system was \textit{``not very well''}~(44\%) or \textit{``not well at all''}~(16\%). Every third participant~(36\%) said their understanding was at least \textit{``somewhat well''}. This is worse than how well they understood why certain posts are recommended by Facebook's News Feed algorithm. For the News Feed, the majority~(56\%) self-assessed their understanding as \textit{``not very well''}~(48\%) or \textit{``not well at all''}~(8\%). This means that the three explanations did not have a measurable effect on the self-reported understanding of users. We also found no difference between those who were able to interact with the systems and those who were not. In the following, we compare the answers of the journalists in our study to the U.S. citizens surveyed by Pew Research Institute~\cite{pewresearch_many_2019}. The majority of U.S. citizens~(53\%) regarded their understanding of Facebook's News Feed as \textit{``not very well''}~(33\%) or \textit{``not well at all''}~(20\%). A larger fraction of U.S. adults thought that their understanding of News Feed is \textit{``somewhat well''}~(32\%). 14\% regarded their understanding of the News Feed as \textit{``very well''}. This implies that the explanations in our investigation did not improve how well participants understood the system and did not improve algorithmic transparency~(RQ1).

\begin{table}[t]
  \centering
  \caption{The three explanations did not help participants understand the personalized curation systems in Study~I. Participants rated the helpfulness from 0~(very little) to 10~(very much).}~\label{tab:helpfulness_ratings}
  \begin{tabular}{|l|c|c|c|c|c|c|}
\hline
& \multicolumn{3}{|c|}{Static} & \multicolumn{3}{|c|}{Interactive} \\
\hline
{Helpfulness} & {$\overline{X}$} & {$\sigma$} & {Mdn} & {$\overline{X}$}  & {$\sigma$} & {Mdn} \\
\hline
System Predictions & \textbf{4.67} & 2.77 & 4.5 & 3.54 & 1.90 & 3.0 \\
Performance Metrics & 2.67 & 1.67 & 2.0 & 3.62 & 2.18 & 4.0 \\
Keywords & 3.50 & 2.91 & 3.0 & \textbf{3.85} & 2.30 & 4.0 \\
\hline
\end{tabular}
\end{table}

Next, we review how the helpfulness of the explanations is perceived by the participants. Those who interacted with the keywords rated performance metrics like accuracy, precision, and recall as the least helpful~(with an average rating of 2.67). System predictions, i.e. seeing the correct predictions as well as false positives and false negatives, were rated as most helpful~(4.67). The keywords received an average rating of 3.50. Those who did not interact with the system rated the system predictions as least helpful~(3.54) and the keywords as the most helpful~(3.85). The performance metrics were rated as 3.62. All of these ratings are below the neutral condition of 5, which indicates that the helpfulness of all three explanations is perceived as low. We found no significant statistical differences between the explanations as measured by the Mann–Whitney U tests, which means that the differences between the ratings could be due to chance. We also found that the ability to interact with the system had no measurable effect. This means that none of the explanations were considered helpful by our participants~(RQ2).

\begin{table}[t]
\centering
\caption{The table shows that participants changing the influence of keywords~(interactive) led to worse system performance.}~\label{tab:interactivity_and_system_performance}
\begin{tabular}{|l|r|r|r|r|r|r|}
\hline
& \multicolumn{2}{|c|}{Accuracy} & \multicolumn{2}{|c|}{Precision} & \multicolumn{2}{|c|}{Recall} \\
\hline
{System}  &  {$\overline{X}$}  & {$\sigma$} &  {$\overline{X}$}  & {$\sigma$} &  {$\overline{X}$}  & {$\sigma$} \\
\hline
Static  & \textbf{78.71} & 7.89 & \textbf{75.53} & 18.43 & \textbf{77.17} & 26.66 \\
Interactive  & 65.87 & 18.02 & 53.09 & 30.66 & 62.00 & 39.49 \\
\hline
\end{tabular}
\end{table}

Table~\ref{tab:interactivity_and_system_performance} shows that curation systems where participants changed the importance of keywords performed considerably worse than those where they did not (RQ3). Personalized ML-based curation systems without participant keywords have 12.84\% better accuracy, 15.17\% higher recall, and 22.44\% better precision. This comparison is based on 5-fold cross-validation. Our in-depth analysis showed that interactive systems for which participants changed a small number of keywords expressing interest performed much better than systems trained by participants that assigned a large number of keywords expressing a lack of interest. One possible explanation for this could be that the keywords selected by participants are not suited to guide ML systems in capturing participants' interests. This is especially surprising considering the framing of the interaction. Participants were not able to freely choose keywords. They only reranked the keywords proposed by the curation system. Nevertheless, the changes they made led to worse system performance. This suggests that the keywords selected by the participants have detrimental effects on the prediction performance of the systems.

\section{Discussion}

We studied explanations in the context of algorithmic news curation. This means that our findings are particularly relevant for those who want to apply ML to recommend news or other content like books, songs, or videos. We found no difference between simple, intuitive, and interactive explanations. None of the three explanations were perceived as helpful by the expert users. Only the intuitive explanation that showed system predictions was rated close to the neutral condition of 5 on the 11-point rating scale. This could imply that the best way to explain an ML-based curation system would be showing the system predictions. This, however, would have some important disadvantages. Unlike ML metrics like accuracy, precision, and recall~(simplicity), or the most influential keywords~(interactivity), it is hard to compare two systems based on their predictions~(intuitiveness). Moreover, the goal of news curation and other ML systems is automation. Evaluating systems by reviewing individual predictions requires a significant time investment. This means that even though system predictions are the most highly rated, they are the least practical of the explanations that we considered. One possible explanation for their appeal is that in contrast to the performance metrics and the influential keywords, the system predictions are directly interpretable and easy to understand. Correct predictions, false positives, and false negatives are straightforward to understand. Overall, our results imply that common strategies of exposing ML systems focused on accuracy, precision, and recall~(simplicity) or the most influential keywords~(interactivity) could be an overextension for users. We, therefore, conclude that intuitiveness is the best paradigm of the three that we tested, even though it was not rated highly in absolute terms. Further research is needed to corroborate this, but considering our highly educated sample of expert users who are familiar with the curation task, it would be surprising if less experienced users benefit from the more complex explanations. 

The key takeaway of the paper is that none of the three explanations were provided as helpful. When users were able to interact with the systems, the performance of the system was much worse. This could imply that the keywords that are important to participants are not the keywords that are important for the curation system. This poses important challenges regarding the direct manipulation of ML-based curation systems and might limit the possibilities for the interaction with curation systems. This is especially problematic because the Gaussian Na\"ive Bayes classifier used in this investigation is a straightforward application of conditional probability, which means that the poor performance is not merely a limitation of this specific classifier. Our findings extend to other statistical machine learning classifiers based on conditional probability because they show that the mathematically important words do not correspond to the words that the user considered to be most important.

Our findings imply that the three approaches to expose curation systems are misguided and need to be reconsidered. None of the three explanations are perceived as helpful by our expert users. The explanations did not improve participants' understanding of the curation system. More than half of the participants said their understanding of the system is “not very well'' (33\%) or “not well at all'' (20\%). This is comparable to how well they think they understand Facebook's News Feed and how well Facebook's News Feed is understood by the average U.S. citizen~\cite{pewresearch_many_2019}. This implies that the explanations did not improve understanding.

Our results indicate a lack of coincidence between the information that can be extracted from a curation system and the information that is meaningful to users. Based on these findings, we introduce the \textbf{Explanatory Gap in Machine Learning-based Curation Systems} to describe the gap between what is available to explain curation systems and what users need to understand such systems. This has important implications for a large body of research on how to explain ML systems~\cite{STUMPF2009639,kim2015interactive,Ribeiro_2016_WIT}. The Explanatory Gap in Machine Learning-based Curation Systems connects to research on the semantic gap in multimedia~\cite{Smeulders:2000:CIR:357871.357873} and the social-technical gap, which Ackerman defined as \textit{``the great divide between what we know we must support socially and what we can support technically''}~\cite{Ackerman:2000:ICC:1463015.1463020}. While the socio-technical gap concerns the lack of technical mechanisms to support the social world, we identified a similar gap regarding the lack of technical mechanisms to support individuals that face complex algorithmic systems. Like the social-technical gap, the Explanatory Gap in Machine Learning-based Curation Systems is unlikely to go away. It is a conceptual framing that can encourage researchers to better understand what is available to explain curation systems and what is needed by users. We hope to encourage further research on how to approach and manage this gap. The finding extends on prior research, e.g., by Rader et al. (2018) \cite{rader_explanations_2018}, who showed that their explanations did not help users evaluate the correctness of a system’s output. However, Rader et al. found that the explanations can make users more aware of how an ML-based system works and that these explanations helped users detect biases. These findings are corroborated by our findings. The findings imply that explanations need to be very simple and easy to understand. Considering the complexity of ML systems, how to achieve this remains an important open question.

This paper is limited by two factors in particular. The professional experience of expert users like journalists could have shaped their perception of how news curation should work and what explanations they consider as helpful. While this potentially limits the generalizability of our findings, if expert users who are familiar with the task of news curation do not benefit from explanations, it is unlikely that users without this background will be able to benefit from the explanations. Our findings are also limited by the high level of education of our participants. The large majority of participants had a university degree~(84\%). However, if even this highly educated subset of the population did not understand these explanations, less educated participants are unlikely to understand them better. Furthermore, we compared our participants' understanding of Facebook's News Feed to a nationally representative sample of U.S. citizens~\cite{pewresearch_many_2019} and found that our findings are generalizable beyond the expert users.

\section{Conclusion}

In this paper, we introduce the Explanatory Gap in Machine Learning-based Curation Systems which describes the gap between what is available to explain ML-based curation systems and what users need to understand such systems. To improve users' understanding of curation systems and to inform algorithmic transparency research, we need further research that explores how such systems should be exposed to users and how the predictions of the systems can be explained. We hope to motivate further experimental studies that explore explanations with real-world tasks like news curation. Future work could investigate how the helpfulness of such explanations is perceived when they are used over a long period, e.g., days, months, or years. The findings indicate that explanations like the most important keywords and interactivity could be an overextension for users. Further research on how well users can understand machine learning systems and, by extension, statistics, would be beneficial. We propose conducting within-subject studies to advance ML explanations and algorithmic transparency. In addition to that, qualitative investigations are needed to explore why the explanations are not perceived as helpful by users. Explorative design studies will be crucial to examine what kind of explanations can help users understand ML-based curation systems.

%
%
%
%
\bibliographystyle{splncs04}
\bibliography{references}
\end{document}